\documentclass[aps, prd, twocolumn, lengthcheck, superscriptaddress, 
nofootinbib]{revtex4-1}

\usepackage{epsfig}
\usepackage[usenames]{color}
\usepackage{graphicx}
\usepackage{amsmath}
\usepackage{multirow}
\usepackage{epstopdf}

\newcommand\sect[1]{\emph{#1.}---}
\def\bi{\bibitem}

\def\la{\langle}\def\ra{\rangle}
\def\be{\begin{eqnarray}}\def\ee{\end{eqnarray}}
\def\lsim{\mathrel{\rlap{\lower3pt\hbox{\hskip1pt$\sim$}}
     \raise1pt\hbox{$<$}}} 
\def\gsim{\mathrel{\rlap{\lower3pt\hbox{\hskip1pt$\sim$}}
     \raise1pt\hbox{$>$}}} 
\def\del{\partial}

\allowdisplaybreaks

\begin{document}

\title{The sound speed and core  of massive compact stars:\\
 A manifestation of hadron-quark duality}

\author{Yong-Liang Ma}
\email{ylma@ucas.ac.cn}
\affiliation{School of Fundamental Physics and Mathematical Sciences,
Hangzhou Institute for Advanced Study, UCAS, Hangzhou, 310024, China}
\affiliation{International Centre for Theoretical Physics Asia-Pacific, Beijing/Hangzhou, China }

\author{Mannque Rho}
\email{mannque.rho@ipht.fr}
\affiliation{Universit\'e Paris-Saclay, CNRS, CEA, Institut de Physique Th\'eorique, 91191, Gif-sur-Yvette, France }

\date{\today}
\begin{abstract}
{
When baryon-quark continuity is formulated in terms of a topology change without invoking ``explicit " QCD degrees of freedom at a density higher than twice the nuclear matter density $n_0$ the core of massive compact stars can be described in terms of fractionally charged  particles, behaving neither like pure baryons nor deconfined quarks. Hidden symmetries, both local gauge and pseudo-conformal (or broken scale),  lead to the pseudo-conformal (PC) sound velocity $v_{pcs}^2/c^2\approx 1/3$ at $\gsim 3n_0$ in compact stars. We argue these symmetries are ``emergent" from strong nuclear correlations and conjecture that they reflect hidden symmetries in QCD proper exposed by nuclear correlations. We establish a possible link between the quenching of $g_A$ in superallowed Gamow-Teller transitions in nuclei and the precocious onset at $n\gsim 3n_0$ of the PC sound velocity predicted at the dilaton limit fixed point. We propose that bringing in explicit quark degrees of freedom as is done in terms of the ``quarkyonic" and other hybrid hadron-quark structure and our topology-change strategy represent the ``hadron-quark duality" formulated in terms of the Cheshire-Cat mechanism~\cite{CC} for the smooth cross-over between hadrons and quarks.  Confrontation with currently available experimental observations is discussed to support this notion.
}
\end{abstract}

\maketitle

\section{ Introduction}
The phase structure of the strong interactions at high densities has been investigated for several decades but still remains very little understood. The observation of massive neutron stars and detection of gravitational waves from neutron star merger provide indirect information of nuclear matter at low temperature and high density, say,  up to  ten times the saturation density $n_0$. So far, such phenomena can be accessed by neither terrestrial experiments nor lattice simulation. Thus the state of baryonic matter of massive compact stars, the densest object in the Universe stable against gravitation collapse, is a totally uncharted domain. 

The study of dense matter in the literature has largely relied on either phenomenological approaches anchored on density functionals or effective field theoretical models implemented with certain QCD symmetries, constructed in terms of set of relevant degrees of freedom appropriate for the cutoff chosen for the effective field theory (EFT),  such as  baryons and  pions, and  with~\cite{baymetal,alford,fukushima-kojo,quarkyonic,lattimer-quarkyonic,margueron} or without~\cite{Holt:2014hma}  hybridization with quarks, including other massive degrees of freedom.  
The astrophysical observations indicate that the density probed in the interior of neutron stars could be as high as $\sim 10$ times the normal nuclear matter density $n_0\simeq 0.16$ fm$^{-3}$ and immediately raise the question as to what the interior of the densest stable star could consist of, say, baryons and/or quarks  and a combination thereof and other stuff that defy the categorization of particles, nuclei etc. Asymptotic freedom of QCD implies that at some superhigh  density, the matter could very well be populated by deconfined quarks as first argued in \cite{collins-perry}. But the density of the interior of stars is far from the asymptotic and hence perturbative QCD cannot be reliable there. Lacking lattice QCD,  high density at low temperature cannot be theoretically accessed.

In this note, a conceptually novel approach going beyond the standard chiral EFT (denoted as $s\chi$EFT) to higher densities $n\gg n_0$ without invoking explicit QCD degrees of freedom is formulated and is used to predict that the core of massive compact stars constitutes of {\it confined} quasi-fermions of fractional baryon charge, encoding the equation of state (EOS) with  the ``pseudo-confomal (PC)" sound velocity (SV) $v_{pcs}^2/c^2\approx 1/3$.  We suggest that this phenomenon can be described by the emergence of certain  symmetries, not explicitly visible in the vacuum, arising from hitherto unexplored nuclear dynamics. Our strategy is to use the characteristics of topology of dense matter consisting of multi-baryons and exploit the topology change in ascending in density to model the baryon-quark continuity believed to be intrinsic in QCD. 
This strategy of exploiting topology to arrive at highly correlated nuclear interactions~\cite{mapping} has an analog in fractional quantum Hall effects where the topology encoded in Chern-Simons field theory -- which is macroscopic -- is mapped to highly correlated electronic interactions in Kohn-Sham density functional theory -- which is microscopic~\cite{jain}. 

Details are given in the review articles written up recently~\cite{MR-PPNP,MR-WS,MR-AAPPS}. Here we will give only  brief description of the basic premise of the theory referring for details to the reviews and focus on the new results obtained and their possible impacts on nuclear astrophysics.

\section{Hidden Symmetries}
There are two essential ingredients that play an indispensable role in our approach. One of them is the topology change that we will return to below. The other is the hidden symmetries we assume to be involved in nuclear dynamics under normal as well as extreme conditions.

In generating higher-density scales for going beyond the  $s\chi$EFT --- that contains only baryons and (pseudo-)Nambu-Goldstone (NG) bosons, we incorporate  two symmetries that are invisible in the vacuum of  QCD: Hidden gauge symmetry and hidden scale symmetry. Both of them could be generated at low density, at higher-order expansions in $s\chi$EFT.  But they must be constrained, particularly at high densities, by the assumed symmetry structure. This is because among others the ultraviolet completion of the effective theory involved is  not available. Our approach is to exploit the possible emergence of these symmetries as density increases to the regime relevant to compact stars, say, $\lsim 10 n_0$.

The higher-energy degrees of freedom we will focus on are the lowest-lying vector mesons $V=(\rho,\omega)$ and the scalar $f_0(500)$.

For the vectors $V$, we adopt the strategy of hidden local symmetry (HLS)~\cite{yamawaki} which at low density is gauge equivalent to nonlinear sigma model, the basis of $s\chi$EFT. This is for two reasons. First, at higher densities, we can assume HLS satisfies Suzuki's theorem~\cite{suzuki}  that states ``when a gauge-invariant local field theory is written in terms of matter fields alone, a composite gauge boson or bosons must be inevitably formed dynamically."  Secondly, the HLS fields could be (Seiberg-)dual to the gluons~\cite{Komargodski,abel,kanetal}. The first accounts for the presence of the ``vector manifestation (VM)"~\cite{VM} that the $V$ meson mass goes to zero as the gauge coupling $g_V$ flows to zero at some theoretically unknown high density $n_{\rm VM}$. We will see below that $n_{\rm VM}\gsim 25 n_0$ is indicated for the emergence of the PC velocity in stars as required by the premise of the composite gauge field structure\footnote{At low density, the composite gauge boson structure is not obvious but if chiral symmetry is the correct symmetry in nuclear dynamics, then HLS being gauge-equivalent to the non-linear sigma model guarantees that higher-order terms of $s\chi$EFT are consistent with the Suzuki theorem.}. Second, it can account for why HLS at the leading power counting works so well and indicate a Higgs phase-topological phase transition coinciding with the quark deconfinement most likely irrelevant to the compact stars we are concerned with~\cite{MR-PPNP}.

Now as for the scalar $f_0(500)$ that we will identify with the dilaton $\chi$ as the NG boson of scale symmetry, we adopt the ``genuine dilaton" (GD) structure proposed in~\cite{crewther}. The key premise of this idea is the existence of a nonperturbative infrared fixed point (IRFP) with the beta function $\beta (\alpha_{s}^{IR})=0$ for flavor number $N_f=3$. The characteristic feature of this IRFP which plays a crucial role in our development is that in the chiral limit both chiral and scale symmetries are in the NG mode. Given the proximity in free-space mass between the kaon and $f_0$, the s(trangeness) flavor figures together with the u(p) and d(own) flavors. In dense medium, however, the effective mass of the s-quark-baryons is to drop less importantly than the light (u,d) quark baryons, thus projecting the GD idea to the two flavors ignoring strangeness as we do in what follows could be justified.


We exploit two versions of the same effective Lagrangian, one purely bosonic incorporating the pseudo-scalar NG bosons $\pi\in SU(3)$, the dilaton $\chi$ and the vectors $V$ in consistency with the symmetries concerned.  We denote the resulting scale-symmetric HLS Lagrangian as $\chi$mHLS with ``m" standing for the meson fields. To do nuclear physics with it, baryons should be generated as skyrmions with $\chi$HLS. This is in principle doable and should provide a nuclear effective field theory approach closest in spirit to QCD given that the skyrmion description for baryons is equivalent to  QCD at least justified in the large $N_c$ limit. But such an approach has defied -- and is defying still -- the efforts to reliably describe normal nuclear matter, not to mention dense matter. An alternative that we resort to is  to introduce baryon fields explicitly into $\chi$HLS, coupled scale-hidden-local symmetrically to $\pi$ and $\chi$. This Lagrangian will be denoted as $\chi$bHLS with ``b" standing for baryons.

\section{Hadron-Quark Continuity and Topology Change}

As will be briefly summarized below (details in \cite{MR-PPNP}),  the $\chi$bHLS applied to baryonic matter  is found to describe nuclear matter at $n\sim n_0$ as well as the currently successful $s\chi$EFT to typically N$^4$LO. There is a however a good reason to believe that the $s\chi$EFT with the Fermi momentum $k_F$ taken as a small expansion parameter must necessarily breakdown at high densities relevant to massive compact stars.  The appearance of quark degrees of freedom in one form or other is a natural candidate for the breakdown mechanism.  In our approach, where and how this breakdown occurs can be determined by the skyrmion approach -- via skyrmion-to-half-skyrmion transition -- with the bosonic Lagrangian $\chi$mHLS~\cite{MR-PPNP}. This is because topology in $\chi$mHLS can be traded in for  what corresponds to hadron-to-quark continuity.

The skyrmion approach, while too daunting to handle nuclear dynamics directly and systematically, can however provide powerful and robust information on the possible topological structure involved in going beyond the normal nuclear matter density regime. At a density  denoted as $n_{1/2} (> n_0)$,  the matter described in terms of skyrmions is found to transform to that of half-skyrmions. This transition has several highly correlated important impacts on the EOS for densities $n > n_{1/2}$. The most crucial of them are that at $n_{1/2}$ (a) the quark condensate $\la\bar{q}q\ra$ vanishes {\it globally} but not locally with non-vanishing pion and dilaton decay constants ($f_\pi\sim f_\chi \neq 0)$, resembling the pseudo-gap phenomenon in condensed matter,   (b) a parity-doubling emerges in the baryon spectrum with the chiral invariant mass $m_0 \simeq (0.6-0.9)m_N$ and (c) the hidden gauge coupling associated with the $\rho$ meson coupling -- which is constant up to $n_{1/2}$ -- starts to drop and flows to zero at the vector manifestation (VM)  fixed point~\cite{VM}, hence the vector meson becoming massless. This is in line with the Suzuki theorem~\cite{suzuki}.

Our key strategy is to import the characteristics of,   and map the robust features of,    topology in $\chi$HLS to the {\it parameters} of the $\chi$bHLS Lagrangian and formulate an EFT going beyond the $s\chi$EFT at density $n> n_{1/2}$ thereby encoding the changeover -- which is not a phase transition --at $n_{1/2}$.  We call this extended nuclear EFT ``generalized nuclear EFT" (acronym $Gen$EFT).  Thus the topology change encoded in the skyrmion structure in $\chi$HLS will be encoded in the parameters of $\chi$bHLS. Thus the putative baryon-quark continuity captured by the topology change at the half-skyrmion density $n_{1/2}$ will involve only one (unique) Lagrangian applicable for the whole range of density from normal nuclear matter density $\sim n_0$ to high density at $\sim (5-7)n_0$ relevant to compact stars.  With the ``continuity" captured solely in the parameters of a Lagrangian, the question is then what are the effective fields before and after the transition at $n_{1/2}$?

To address this question, it is helpful to note that the skyrmion-to-half-skyrmion transition in the skyrmion matter has an analog -- though physics is vastly different -- in condensed matter in (2+1) dimensions, i.e.,  the transition from the magnetic N\'eel ground state to the VBS (valence bond solid) quantum paramagnet phase~\cite{DQCP}.\footnote{Recent development indicates this elegant picture of DQC~\cite{DQCP}  may no longer be realistic~\cite{sachdev}.  How the basic idea is evolving in this phenomenon is a warning  to our work that one cannot be too naive in subtle quantum critical phenomena and what we are doing in this strong interaction case could  be misleading.}   The half-skyrmions intervening in this case are {\it deconfined}, thus the transition involves ``deconfined quantum critical (DQC) points" but with no local order parameters for the phase transition.  What takes place in dense skyrmion matter which in some sense resembles the DQC phenomenon with no order parameter seems, however, quite different because the half-skyrmions are not deconfined but confined by monopoles~\cite{cho}. This suggests that the confined half-skyrmion ``complex" could be treated, although the symmetry is quite different, as a {\it local} baryon number-1 field with its parameters, such as the mass, coupling constants etc. drastically modified from the vacuum quantities reflecting the topology of the half-skyrmions. 
\section{$Gen$EFT}

Given the $\chi$bHLS Lagrangian with the characteristics described above, how does one go about formulating $Gen$EFT, that is, doing nuclear many-body problem ?

With the $\chi$bHLS Lagrangian whose parameters are suitably fixed in medium matched to QCD correlators~\cite{VM} and mapped from the skyrmion topology,  it should in principle be feasible to do systematic nuclear scale-chiral perturbation calculation with $\chi$bHLS, suitably constructing a scale-chiral counting rule including the hidden gauge fields and dilaton field. Unfortunately, as formulated up-to-date, there are too many unknown parameters~\cite{MR-PPNP} so that high-order calculations have not been studied yet. Instead we resort to formulating many-body nuclear problems by a Wilsonian renormalization-group (RG) type approach to arrive at Landau(-Migdal) Fermi-liquid theory. The mean field approximation with $\chi$bHLS can be identified with Landau Fermi-liquid fixed point (FLFP) theory valid in the limit $1/\bar{N}\to 0$ where $\bar{N}=k_F/(\Lambda-k_F)\sim E_F/E$ (with $\Lambda$ the cutoff on top of the Fermi sea)~\cite{benfatto,polchinski,shankar}. This approach can be  taken as a generalization of the energy-density functional theory familiar in nuclear physics~\cite{MR91,MR-PPNP}. One can go beyond the FLFP in  $V_{lowk}$-RG as we will explain below.
\subsection{Nuclear Properties at Low Density}
\subsubsection{Normal nuclear matter}
When the $Gen$EFT so formulated is applied to the usual nuclear matter at density $\sim n_0$, it is found to work fairly well, at least as well as the s$\chi$EFT up to, say, N$^4$LO~\cite{MR-PPNP}. It is likely to work up to $\sim 2n_0$ as in s$\chi$EFT. With the heavy vector and scalar degrees of freedom explicitly figuring in the dynamics at the tree order with the cutoff set above the mass of the heavy DoFs, the power counting rule of course gets modified from that of s$\chi$EFT. The mean field with $\chi$bHLS with the parameter scaling with density captured in the dilaton condensate $\la\chi\ra$ and in the FLFP approximation  could well be comparable to the high-order s$\chi$EFT calculation. Furthermore --- and this is not recognized in nuclear circles --- there is a magical element in HLS, as stressed in \cite{Komargodski,kanetal,abel}, possibly connected, as conjectured, with the Seiberg-duality between HLS and gluons of QCD.

Just to give an idea, we quote the predictions of a few thermodynamic properties of nuclear matter obtained for $n_{1/2}=2.5n_0$ compared in parentheses with available empirical information: $n_0 = 0.161 (0.16\pm 0.01)$ fm$^{-3}$,  B.E. $= 16.7 (16.0\pm 1.0)$ MeV, $E_{sym} (n_0)=30.2 (31.7\pm 3.2)$ MeV,  $E_{sym} (2n_0)= 56.4 (46.9\pm 10.1; 40.2\pm 12.8)$ MeV, $L(n_0)=67.8 (58.9\pm 16;  58.7\pm 28.1)$ MeV, $K_0=250.0 (230\pm 20)$ MeV. Obtained with little fine-tuning in parameters, the results are as good as in high-order s$\chi$EFT.
\subsubsection{``Quenched $g_A$" in nuclei in Fermi-liquid fixed point theory}
That the standard high-order $s\chi$EFT and the $Gen$EFT with the DoFs encoding hidden symmetries fare equally well at $n\sim n_0$ indicate that those symmetries are buried and not apparent in the EOS at that density. An illuminating case for this indication is the well-known puzzle in nuclear Gamow-Teller beta decay transitions where the simple shell-model description requires that the ``effective Gamow-Teller coupling constant" be $g_A^\ast\approx 1$, quenched from the neutron decay value $g_A=1.276$. 

This puzzle has a long history dating from 1970's. We will highlight the essential points directly relevant to the principal thesis we will develop below in what we consider to be the {\it resolution} of the $g_A$ puzzle as described in \cite{gA}. More details and relevant references are given in the Letter paper.

Consider a superallowed Gamow-Teller (GT) transition in the extreme single-particle shell model (ESPSM for short) from a doubly magic nucleus. We pick in particular the nucleus $^{100}$Sn~\cite{sn100,RHIC}.\footnote{In this case,  the proton in a completely filled shell $g_{9/2}$  beta-decays via the spin-isospin flip to a neutron in an empty shell $g_{7/2}$ at zero momentum transfer.}  Let us denote the matrix element so calculated  as ${\cal M}_{\rm ESPSM}$.  Suppose that the transition matrix element is accurately measured and we denote it ${\cal M}_{\rm exp}$. The ``quenching factor $q$" is defined by
\be
{\cal M}_{\rm exp}= q {{\cal M}_{\rm ESPSM}}.\label{quenching}
\ee
Now to see what $q$ stands for, we need to specify the axial current for the GT transition. The axial current in the $\chi$bHLS Lagrangian figuring in $Gen$EFT is given by~\cite{gA} 
\be
J_{A\mu}^a=\kappa(\chi)g_A\bar{\psi}\gamma_\mu\gamma_5\tau^{\pm}\psi\label{Acurrent}
\ee
where
\be
\kappa(\chi)=c_A +(1-c_A)(\frac{\chi}{f_\chi})^{\beta^\prime},\label{delta}
\ee
$\beta^\prime$ is the anomalous dimension of the gluon stress tensor and $c_A$ is a constant, undetermined in the theory, $\chi$ is the conformal compensator which we take for the dilaton field\footnote{The $\chi$ field, transforming linearly under scale transformation, is related to $\sigma$ which transforms nonlinearly as $\chi=f_\chi e^{\sigma/f_\chi}$. This is like the chiral field $U=e^{i\pi/f_\pi}$.},  and $f_\chi$ is the dilaton decay constant. Note that the axial current (\ref{Acurrent}) has both the scale-invariant component,  the first term of (\ref{delta}) and the scale-breaking component,  the second term of (\ref{delta}). The current is scale-invariant only when $c_A=1$ (for which $\kappa(\chi)=1$) given that we have $\beta^\prime\neq 0$~\cite{crewther}.

In medium, $\chi$ picks up a ``vacuum" (medium) expectation value (VEV), so (\ref{delta}) will have a term that contains $f_\chi^\ast/f_\chi$ where the asterisk stands for density dependence and a term that contains the fluctuating dilaton fields. The latter contributes only at loop orders to the GT matrix elements, so we can drop the fluctuating terms.  We define
\be
d_{ssb}\equiv \kappa(\la\chi\ra^\ast)=c_A +(1-c_A)\Phi^{\beta^\prime}
\ee
where~\cite{BR91}\footnote{This equality holds in medium at the chiral matching scale $\Lambda_\chi\sim 1$ GeV.}
\be
\Phi=f_\chi^\ast/f_\chi=  f_\pi^\ast/f_\pi.
\ee

We now consider the matrix element of the scale invariant current $g_A\bar{\psi}\gamma_\mu\gamma_5\tau^{\pm}\psi$. The matrix element ${\cal M}_{\rm ESPSM}$ is given by the transition matrix element taken between the initial ESPSM (extreme single-particle shell-model) state (i) and  the final ESPSM state (f) by the single-particle operator $g_A\tau^\pm \sigma$ that we denote as $g_A(\tau^\pm\sigma)_{fi}$.    

Now write the ``exact" transition matrix element ${\cal M}_{\rm GT}\equiv \la F |g_A\bar{\psi}\gamma_\mu\gamma_5\tau^{\pm}\psi | I  \ra$ where $\la  I (F) |$ is the exact parent (daughter)-state wave function (in shell model)
\be
{\cal M}_{\rm GT} &=& q_{\rm snc} {\cal M}_{\rm ESPSM} = q_{\rm snc} [g_A(\tau^\pm\sigma)_{fi}]\nonumber\\
& =& (1+\Delta) [g_A(\tau^\pm\sigma)_{fi}] \label{MGT}.
\ee
As defined, $q_{\rm snc}$ is to capture the ``complete"  strong nuclear correlations.  Now the single-particle GT operator $\tau^\pm\sigma$ couples strongly  not only to one-particle-hole states but also to multi-particle-multi-hole states of GT quantum numbers by the nuclear tensor forces involving excitation energies up to  $\lsim 300$ MeV. The effect of screening is to be captured intirely in $\Delta$, which is typically negative. In practice,  full {\it no core} shell-model calculations to obtain $q_{\rm scs}$ would be highly daunting if not impossible  in such double-magic nuclei like $^{100}$Sn. To the best of our knowledge, such calculations do not exit up to date.\footnote{There are efforts in lighter double-magic nuclei such as $^{48}$Ca that do clearly indicate that high-order correlations are crucial~\cite{ca48}.}  It thus remains a challenge to a true ``first principles calculation"  combining the s$\chi$EFT, not to mention the $Gen$EFT,  with powerful state-of-the-art quantum many-body techniques.\footnote{ There are also many-body currents that contribute,  but for consistency in power counting in $Gen$EFT for superallowed transitions, it is justified to drop them for the superallowed GT transitions ~\cite{gA}. This is in contrast to the time component of the axial current, e.g., in first-forbidden transitions.}. 

It follows from (\ref{quenching}) and (\ref{MGT}) that
\be
q=q_{ssb}q_{snc}.
\ee
What is referred to as ``quenched $g_A$" observed in light nuclei is~\cite{suhonen}\footnote{Caveat: the GT transitions concerned are not necessarily in ``ESPSM." Hence certain limited correlations are involved in some of the results available in the literature~\cite{suhonen}. }
\be
g_A^\ast=q g_A\approx 1.\label{lightnuclei}
\ee
If one ignores the scale anomaly effect and set $q_{ssb}=1$,  then $g_A^\ast\approx 1$ will be {\it entirely} due to nuclear correlations $q_{snc}$.

The  problem therefore boils down to what $q_{snc}$ is in the ESPSM case. It is here that our $Gen$EFT can make a  powerful contribution. It is not in an all-order calculation in s$\chi$EFT (which  is not feasible at present)  but in the Fermi-liquid fixed point approach sketched above that the solution is found. 

It is the precise mapping of the $q_{\rm snc}$ given in the Fermi-liquid fixed point approach to the $q_{\rm snc}$ defined in the ESPSM. How the mapping is done is as follows.

As alluded above, the basic strategy of $Gen$EFT is to apply the scale-symmetric baryonic ($\chi$bHLS) Lagrangian  to a Wilsonian renormalization-group approach to strongly interacting fermions  on the Fermi surface~\cite{benfatto,polchinski,shankar}. To the best of our knowledge, there is no rigorous derivation, but the chain of our arguments to go from, say, a chiral Lagrangian to the Fermi-liquid structure of many-nucleon systems go as follows. First it has been established that doing Walecka mean field theory~\cite{walecka} with vector mesons (not necessarily HLS)  and a scalar meson  (not necessarily dilaton)  leads to Fermi-liquid model~\cite{matsui}. Doing the mean-field with $\chi$bHLS gives a very good description of normal nuclear matter~\cite{song,MR-PPNP}. It turns out that the mean field $\chi$bHLS  corresponds to the Fermi-liquid fixed-point theory (FLFP)~\cite{MR91}. Going  beyond the fixed point taking into account higher-order corrections in $E/E_F$ has been formulated in $V_{lowk}$ RG~\cite{PKLR,PKLMR} which  corresponds to the double decimation RG while the FLFP theory is of single-decimation RG.

Surprisingly the result for $q_{snc}$ was obtained in a simplified treatment of the $\chi$bHLS even a long time ago~\cite{BR91}. The profound significance of the result, stressed below, was not recognized then. What turns out to correspond to (\ref{MGT}) is given by the matrix element of the single-particle GT operator given by decimating the Wilsonian RG integral all the way down to the Fermi surface in the large $\bar{N}$ limit which corresponds to the Fermi-liquid fixed-point approximation. The superallowed matrix element of the Gamow-Teller transition from the quasi-proton (-neutron) to the quasi-neutron  (-proton) on top of the Fermi sea is then simply given by the  Landau quasiparticle mass $m_N^L $ and $\Phi$ that contains the  dilaton condensate~\cite{BR91,gA} 
\be
q_{snc}=(1-\frac 13\Phi \tilde{F}_1^\pi)^{-3}\label{Landau}
\ee
where $\tilde{F}_1^\pi=\frac{m_N}{m_N^L}F_1^\pi$ with $m_L$ the Landau mass,  and $F_1^\pi$ is the Landau parameter contributed by  the pion exchange which is accurately calculable at densities near $n_0$. This result is basically a low-energy theorem result, specifically a Goldberger-Treiman-type relation, involving the dilaton (in place of the pion),  valid for the large $N_c$ and large $\bar{N}$ limits. Now $\Phi=f_\pi^\ast/f_\pi$is given by deeply bound pionic nuclei (known for $n\leq n_0$). 

It turns out that the quantity $\Phi \tilde{F}_1^\pi$ is a nearly independent of density -- due to cancellation -- for $n \sim (1/2-1)n_0$. So it is more or less the same  in light as well as  heavy nuclei.

Within the theoretical uncertainties in the range of densities involved, we get from Eq.(\ref{Landau})
\be
q_{snc}= 0.78 (1\pm 20\%)
\ee
which leads to what we will call ``Landau $g_A$" 
\be
g_A^{\rm Landau}\equiv g_A q_{snc}= 1.0\pm 0.2.\label{LandaugA}
\ee 
This is  what's found in light nuclei (\ref{lightnuclei})~\cite{suhonen}.

Now the key point of our arguments is that (\ref{LandaugA}) is what should be given by the ESPSM with $g_{\rm ssb}$ set equal to 1. In other words, {\it it should be entirely given by ``nuclear correlations."}

Now the question is: Is (\ref{LandaugA}) just an accident or something more profound?

Interestingly, it has been shown~\cite{bira-dlfp}  that starting with non-linear sigma model with constituent quarks and after certain field redefinitions, letting the dilaton mass $m_\chi$ go to zero with the conformal anomaly turned off --- and in the chiral limit --- one arrives at  a linearized Lagrangian that satisfies various well-established sum rules, such as, among others, the Adler-Weisberger sum rule,  provided the singularities that appear as $m_\chi\to 0$ are suppressed by what are called ``dilaon-limit  fixed-point (DLFP)" constraints. It has been verified that the same limiting process to our $\chi$bHLS yields the constraints of \cite{bira-dlfp}
\be
g_V&\to& g_A\to 1,\label{gagv}\\
f_\pi &\to& f_\chi\neq 0.\label{bira}
\ee
We refer to  the dilaton fixed point $g_A$ as $g_A^{\rm DL}$.

What this implies is that the quasiparticle, whether baryonic or quarkonic or whatever unparticle-ish, tends to carry the axial charge $g_A^\ast=1$ manifesting scale invariance. Thus the ``pseudo-conformal structure" -- given that dilaton mass is not zero -- permeates from $\sim n_0$ to a dilaton-limit-fixed-point density $n_{\rm dlfp}\gg n_0$. Whether this conjecture is supported, at least partially, in nature could be checked in precision experiments in doubly magic nuclei as proposed in \cite{gA}. 

The present experimental situation is not yet clear. Among several experiments available in the literature, let us take two most recent ones in $^{100}$Sn~\cite{sn100,RHIC}.  The RHIC data~\cite{RHIC} has been claimed to be an improvement over the GSI data~\cite{sn100}.  The ranges of $q_{snc}$ extracted from the data -- within the error bars -- are found to be
\be
q_{snc}^{\rm RHIC} &=& 0.55-0.45 \to g_A^\ast = 0.70-0.57 \nonumber\\
q_{snc}^{\rm GSI} &=& 0.82-0.61 \to g_A^\ast =1.05-0.78.
\ee
Given the large error bars of the GSI data, one can say little on the scale anomaly effect $d_{\rm ssb}=1 - g_A^\ast$. The RHIC data, if correct,  do however indicate that there can be anomaly contribution~\cite{gA}. A more precise data would be extremely interesting for both the emergence of scale invariance (\ref{LandaugA}) in nuclear correlations and its possible violation by anomaly which will zero-in on $\beta^\prime$ in nuclear matter~\cite{gA}.

\subsection{Emergent Scale Symmetry in Compact Stars}
We now address what this business of the ``quenched $g_A$" has anything to do with massive neutron stars.

Our suggestion or rather our conjecture is that  it links how this comes about in nuclei to what may be taking place at high density in the core of massive stars. This story was told before, but we find  the potential role of (spontaneously broken) scale symmetry buried in the $g_A$ problem {\it manifested} most strikingly in leading to the pseudo-conformal sound velocity $v_s^2\approx 1/3$ for density $n > n_{1/2}\approx 3n_0$ and the ``stuff" in the core of massive stars. 

\subsubsection{Equation of state  of compact stars}
We now turn to the application of the $Gen$EFT formalism to the structure of massive neutrons stars based on the standard TOV equation. How it fares with the (observed) global properties of massive stars is detailed in a number of reviews (e.g., \cite{MR-PPNP}), so we will focus exclusively on the predictions made in $Gen$EFT that are unique in that they do not seem to be shared by -- and in some cases, drastically different from--  other models in the field. We will also discuss with a dose of speculation on the implication of what we ``find" on  the structure of the densest baryonic matter stable against collapse into black hole.

To apply the $Gen$EFT constructed as described above to compact stars, what  we first need  is to locate the density at which the topology change that we identify with the putative hadron-to-quark continuity takes place. It is fixed by neither theory nor experiments. It turns out however~\cite{MR-PPNP} that the available astrophysics phenomenology does provide the narrow range  $2\lsim n_{1/2}/n_0 \lsim 4$.  In the $Gen$EFT approach, there is little variation in the EOS within this range,  so $n_{1/2}=2.5n_0$ is good enough for illustration.

Up to $n_{1/2}$, the same EOS that works well at $n_0$ is assumed to hold.  It is what comes after $n_{1/2}$ due to the topology change that plays the crucial role for the properties of compact stars.  

The most important quantity from the topology change is the cusp in the symmetry energy in the EoS that results from the interplay between the pion and the heavy degrees of freedom, principally the $\rho$ meson. In $Gen$EFT, this translates into the gauge coupling $g_\rho$ going to zero at the VM fixed point $n_{\rm VM}$. This dropping coupling constant affects the nuclear force, specially the tensor force, such that the EoS, soft below $n_{1/2}$, becomes hard above. This provides the crucial mechanism needed for the observed massive stars.

Equally important is the approach to the dilaton-limit fixed point (DLFP) --  associated with the IR fixed point in the GD scheme -- assumed to be close or equal to the VM fixed point $n_{\rm VM}$. The consequence is that $g_A\to 1$ as mentioned before and the dilaton decay constant $f_\chi\to m_0$, the (density-independent) chiral invariant baryon mass, due to an interplay between the dilaton and the $\omega$ meson~\cite{PKLMR}.  This makes the dilaton decay constant $f_\chi$ {\it albeit roughly}\footnote{This reserved remark is due to a caveat mentioned below.} density-independent in the range between $n_{1/2}$ and the core density of the star.

A few short remarks are in order here for completeness.

The star properties obtained in $Gen$EFT are found to be generally consistent with presently available observations~\cite{MR-PPNP}. For  $n_{1/2}=2.5 n_0$,  the maximum star mass is found to be $M_{max}\sim 2.0 M_\odot$ with the central density $\sim 5n_0$. For a neutron star with mass $1.4M_{\odot}$, currently highly topical in connection with gravity-wave data,  we obtain the dimensionless tidal deformability $\Lambda_{1.4} \approx 656$ and the radius $R_{1.4} \approx 12.8$~km.
Were we to pick $n_{1/2}=4.0 n_0$, the upper bound in the parameter choice,  the maximum star mass would reach $M_{max}\simeq 2.25 M_\odot$.
\footnote{If $\Lambda_{1.4}$ were confirmed to be much lower than 656 as claimed in some publications and/or if the maximum star mass were measured to be higher than, say, 2.3 $M_\odot$, then we must admit that our theory could get into tension with the observations.}

\subsubsection{The role of VM fixed point} 
Our theory makes  two predictions drastically different from those of all  other ``standard" EFTs available in the literature. And they are in the sound velocity (SV) of stars $v_s$ and the  impact on the structure of the core of massive stars.

Before proceeding further, we recall the existence of the VM fixed point is required by the {\it assumption} that the vector mesons are ``composite gauge fields"~\cite{suzuki}.  It turns out that there is a great difference in what we discuss below if one takes $n_{\rm VM}\sim 6 n_0$ or $n_{\rm VM}\gsim 25 n_0$. The former is usually taken in the literature as the density at which ``deconfined  quarks" could appear. Most of the global neutron star properties are more or less insensitive to which value of $n_{\rm VM}$ is taken, but there is a drastic difference in the structure, i.e., the constituents,  of the core of massive stars as well as in the sound velocity.

In what follows we take $n_{\rm VM}\gsim 25 n_0$ and comment on what happens when $n_{\rm VM}\sim 6n_0$ is taken.

We will base our discussion on what has been obtained in \cite{PKLMR}. There $Gen$EFT is formulated in $V_{lowk}$RG formalism which is essentially equivalent to doing Landau Fermi-liquid theory using $\chi$bHLS Lagrangian going beyond the infinite  $\bar{N}$ limit\footnote{Higher-order corrections in $1/\bar{N}$ are approximated by ring diagrams. Relevant references are given in \cite{PKLR,PKLMR}.}. 
\subsubsection{The pseudo-conformal (PC) sound velocity } 
\begin{figure}[h]
\begin{center}
\includegraphics[width=7.7cm]{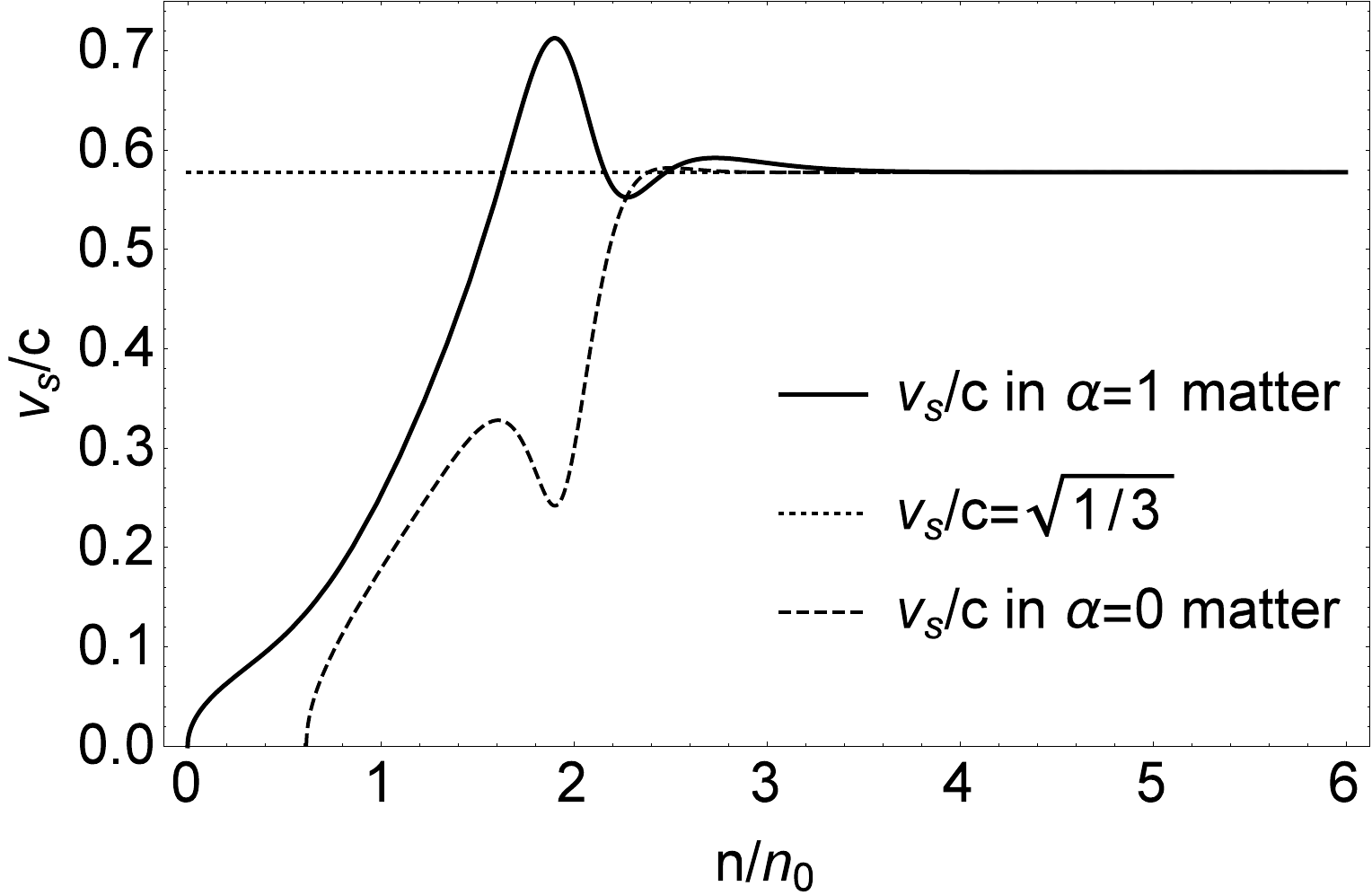}
\caption{ The sound velocity $v_s/c$ vs. density for $\alpha \equiv \frac{N-Z}{N+Z}=0$ (nuclear matter) and $\alpha=1$ (neutron matter) in $V_{lowk}$ RG for $n_{1/2}=2 n_0$ and $v_{\rm VM}=25 n_0$.
 }\label{SV}
 \end{center}
\end{figure}

\begin{figure}[h]
\begin{center}
\includegraphics[width=7.5cm]{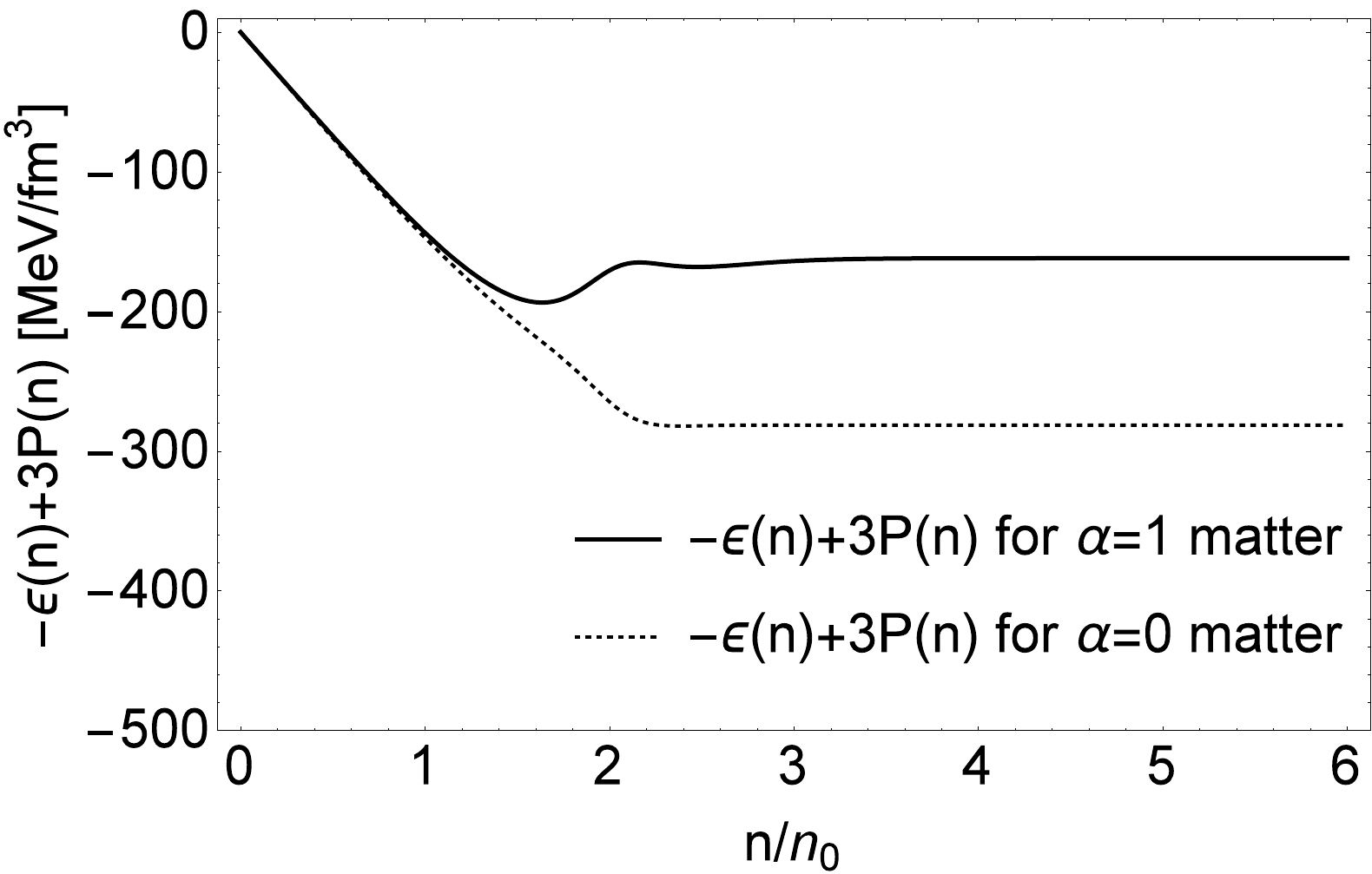}
\caption{$\la\theta_\mu^\mu\ra$  vs. density for $\alpha=0$ (nuclear matter) and $\alpha=1$ (neutron matter) in $V_{lowk}$ RG for $n_{1/2}=2 n_0$ and $v_{\rm VM}=25 n_0$.
 }\label{TEMT}
 \end{center}
\end{figure}

The sound velocity $v_s$ 
\be
v_s=\del P(n)/\del \epsilon (n)
\ee
given by the pressure $P(n)$ and the energy density $\epsilon (n)$ obtained in $Gen$EFT~\cite{PKLMR} is plotted in Fig.~\ref{SV}\footnote{We give  the results for $n_{1/2}=2n_0$. They are essentially the same as for $n_{1/2}=2.5n_0$ that will be used below.}. 
There are two prominent observations to make here. One  is the {\it precocious} convergence to $v_s^2/c^2=1/3$ at $n\sim 3n_0$  and the other is the bump (mainly from neutrons) above $v_s^2/c^2=1/3$ between $n_0$ and $2n_0$ going above 1/3 and then converging to 1/3 from slightly below.  It is easy to see the convergence to $v_s^2/c^2=1/3$ if one looks at the trace of the energy momentum tensor (TEMT) $\la\theta_\mu^\mu\ra$. If  $\la\theta_\mu^\mu\ra=0$ as expected  at the asymptotic density, the ``conformal sound velocity" $v_s^2/c^2=1/3$ is automatic, but the density of stable compact stars is not asymptotic, certainly not  at such low density $n \gsim 3n_0$. To see what is happening, let's look at the behavior of $\la\theta_\mu^\mu\ra$ vs. density. It is plotted in Fig.~\ref{TEMT}.
One notes first that $\la\theta_\mu^\mu\ra\neq 0$  which is natural since the system is away from the IR fixed point, i.e., the dilaton mass is not zero. Second it is, within the approximation involved,  a density-independent constant for both the neutron matter and nuclear matter for $n\gsim n_{1/2}$. Thus
\be
\frac{\del}{\del n} \la\theta_\mu^\mu\ra=\frac{\del\epsilon(n)}{\del n}\left(1-3v_s^2\right)\approx 0.
\ee
As far as we know, there is no Lee-Wick-type state at the densities relevant to massive stars, so $\frac{\del\epsilon(n)}{\del n}\neq 0$.  Therefore  it must  be that
\be
v_s^2/c^2\approx 1/3.\label{PCSV}
\ee
It should be stressed here that given the approximations involved, the exact equality cannot be expected. We therefore call this velocity ``pseudo-conformal (PC) sound velocity."

The presence of lepton degrees of freedom would, of course, be required to address realistically the properties of compact stars. The proton fraction vs. density required has been worked out in \cite{PKLMR}. It does not however affect the prediction of the PC sound speed (\ref{PCSV}) -- apart from the bump region where the changeover from hadronic to non-hadronic takes place and must be quite involved given the same constancy vs. density.

It is significant to note here that this result is in fact expected already at the mean-field level, namely, at the leading order in $1/\bar{N}$ expansion in $Gen$EFT. In the mean field of  $Gen$EFT, the VEV of the trace of energy momentum tensor takes the simple form $
\la\theta^\mu_\mu\ra=4V(\la\chi\ra) -\la\chi\ra\frac{\del V(\chi)}{\del\chi}|_{\chi=\la\chi\ra}$ where $V(\chi)$ is the dilaton potential where the conformal anomaly effect is included. As noted above, $\la\chi\ra\propto f_\chi$ is density-independent in the range of density involved in the interior of compact stars, it therefore follows  $
\frac{\partial}{\partial n}\langle \theta_\mu^\mu \rangle =0.$ 

That this result of the pseudo-conformal structure is given in the Fermi-liquid fixed point approximation and also in the $V_{lowk}$ RG going beyond the large $\bar{N}$ limit suggests that we could simply replace the EoS for $n\geq n_{1/2}$ by a simple analytic form. This would facilitate then varying the topology change density $n_{1/2}$. This turns out to be convenient for addressing the recent gravity-wave observations such as the dimensionless tidal deformability ${\Lambda}_{1.4}$~\cite{MR-PPNP}. This has been explicitly verified for $n_{1/2}= (2-3)n_0$.  The analytic form takes the simple form for $n\geq n_{1/2}$
\be
E/A=-m_N +X^\alpha \big(n/n_0)^{1/3} + Y^\alpha (n/n_0)^{-1}\label{EoverA}
\ee
with $A=N+Z$ and $\alpha =(N-Z)/(N+Z)$.  The constants $X$ and $Y$ depend on $n_{/2}$ and are to be fixed by the continuity of the chemical potential and pressure at $n=n_{1/2}$ with the $E/A$ calculated in $V_{lowk}$RG for $n \lsim n_{1/2}$. It is easy to check that (\ref{EoverA}) gives $v_s^2/c=1/3$ independently of $X$ and $Y$. It is clear that the fine details of the ``bump" in the sound velocity in the vicinity of the topology change density cannot be trusted given the sharp cross-over which is not realistic. It turns out that for the case studied with $n_{1/2}=2.5n_0$, this construction precisely reproduces the $V_{lowk}$ RG result.

\begin{figure}[h]
\begin{center}
\includegraphics[width=7.0cm]{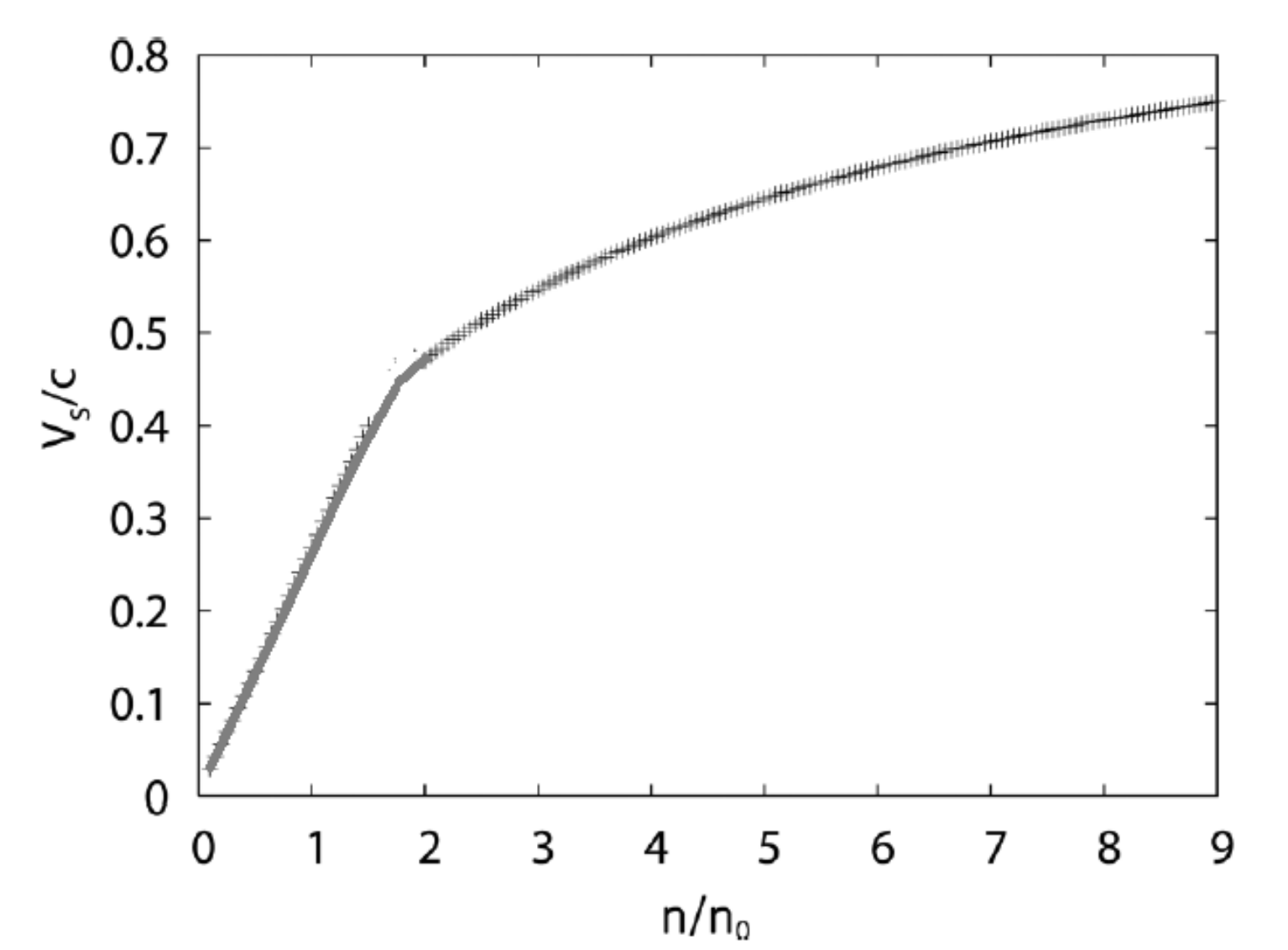}
\caption{ The sound velocity $v_s/c$ vs. density for neutron matter in $V_{lowk}$ RG for $n_{1/2}=2 n_0$ and $n_{\rm VM}=6 n_0$.
 }\label{vm6}
 \end{center}
\end{figure}

In stark contrast  to Fig.~\ref{SV} for $n_{\rm VM}\gsim 25 n_0$, the sound velocity for $n_{\rm VM}\sim 6n_0$,  as one can see in Fig.~\ref{vm6},  keeps increasing continuously above the conformal value at  $n\gsim 3n_0$.  Thus the PC structure given by (\ref{EoverA}) is simply destroyed. Furthermore at higher densities, it can even violate the causality bound $v_s/c=1$. This behavior is quite similar to what one gets in RMF approaches found in the literature. See for instance \cite{RMFT}. If correct, this remarkable feature suggests an extremely intricate interplay  between the VMFP and the emergence of scale symmetry resembling the feature observed in the quenching of $g_A$. Whether this is coincidental or not is not understood. 
If it is not coincidental, it could very well reflect the emergence of hidden scale symmetry from nuclear matter to compact-star matter put forward above.  

We make the following proposition.

In the range of density relevant to the densest stable object in the Universe, far from asymptotic density in QCD, {\it approximate} hidden scale symmetry emerges and becomes effective. Given the inevitable intricate interplay between the dilaton  (attractive)  interactions  and the $\omega$  (repulsive)  interactions in baryonic matter,  it must be that with $m_\chi\neq 0$, the state  of the matter must be away from the IR fixed point measured by the distance of the QCD gauge coupling $\alpha_s$ from the IR fixed point $\alpha_{\rm IR}$ multiplied by the anomalous dimension $\beta^\prime$~\cite{crewther}. This  means that the dense matter system we are dealing with must have the scale symmetry broken by anomaly, so that certain quasiparticle structure is  preserved. If it were at the IR fixed point with the scale symmetry restored, however, the system must be something else than conventional baryonic or quarkonic, perhaps even   ``unstuff"-ish  losing quasiparticle structure~\cite{nonFL,georgi}.

It is interesting to confront the above proposition with a recent analysis that combines astrophysical observations and model independent theoretical {\it ab initio} calculations~\cite{evidence}. Based on the observation that, in the core of the maximally massive stars, $v_s$ approaches the conformal limit $v_s^2/c^2 \to 1/3$ and the polytropic index takes the value $\gamma < 1.75$ --- the value close to the minimal one obtained in hadronic models --- Annala et al. arrive at the ``conclusion" that the core of the massive stars is populated by  what can be considered as deconfined quarks.

\begin{figure}[htbp]
\begin{center}
\includegraphics[width=0.4\textwidth]{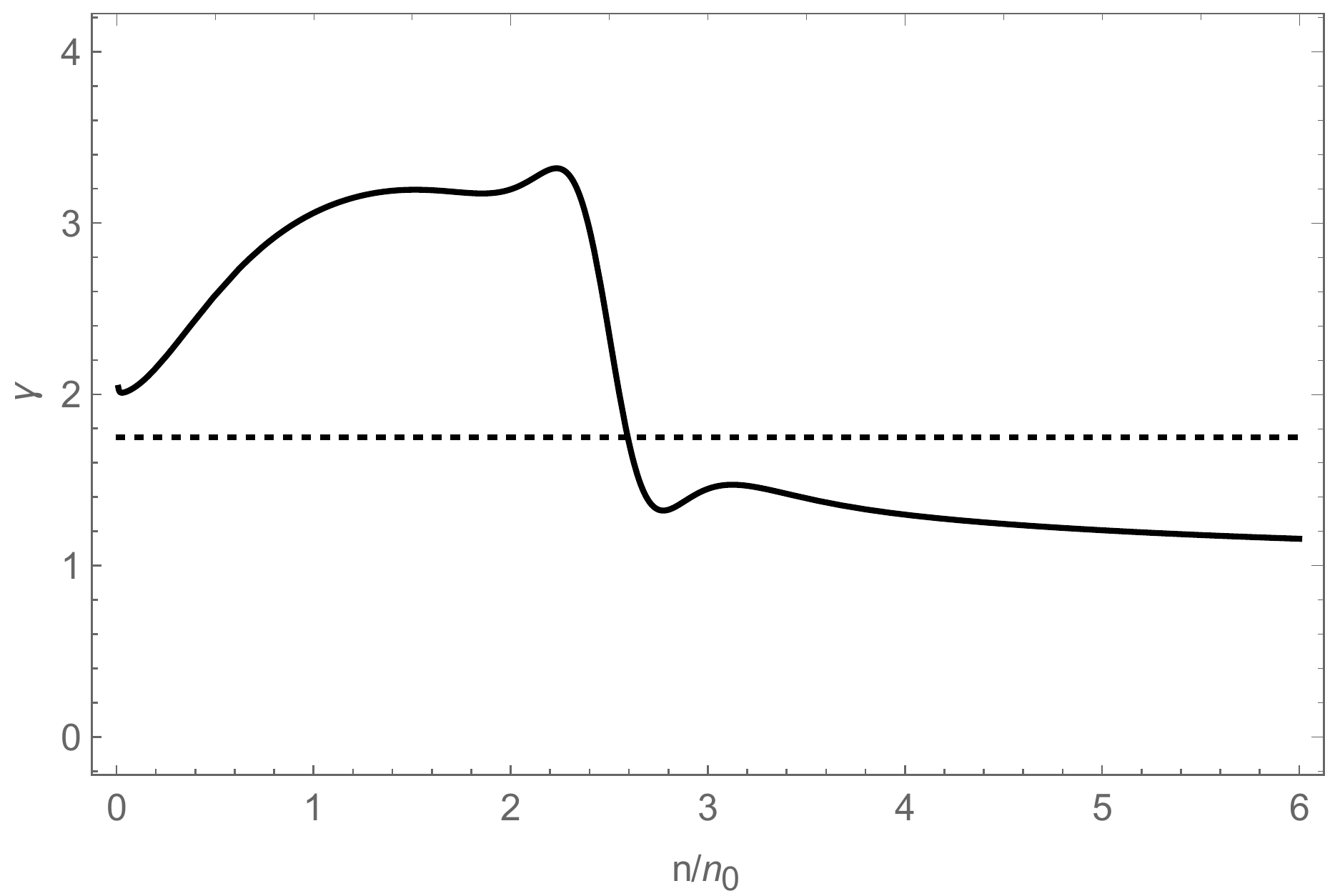}
\end{center}
\vskip -.5cm
\caption{Density dependence of  the polytropic index  in neutron matter.}
\label{polytropic}
\end{figure}

\begin{figure}[bhtp]
\begin{center}
\vskip 0.3cm
\includegraphics[width=0.4\textwidth]{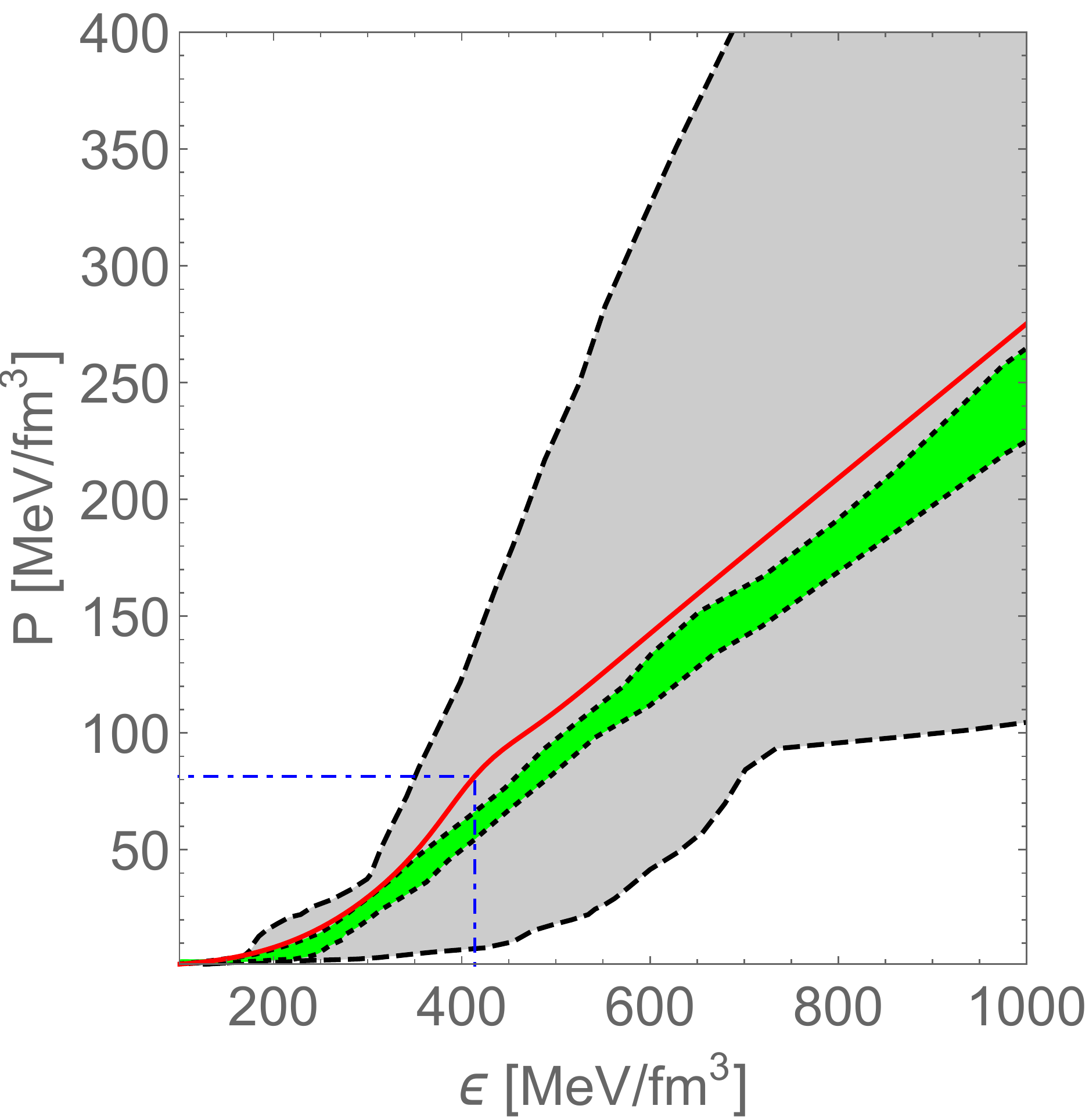}
\end{center}
\vskip -0.5cm
\caption{Comparison of $(P/\epsilon)$  between the PCM  velocity and the band generated with the
SV interpolation method used in~\cite{evidence}. The gray band is  from the causality and the green band from the conformality. The red line is the PCM prediction. The dash-dotted line indicates the location of the topology change.}
\label{fig:EoS}
\end{figure}

Our PC structure has the PC sound velocity  $v_{pcs}^2/c^2\approx 1/3$ for $n\gsim 3n_0$, hence in the core of the star. Furthermore the polytropic index defined by
\be
\gamma= {d\ln P}/{d\ln \epsilon}
\ee
plotted in Fig.~\ref{polytropic}
shows a large $\gamma\sim 3$ below $n_{1/2}$ expected in nuclear matter,  drops below 1.75 at the topology change and then goes to near 1 at the core density $\sim 5 n_0$ of the star.  Thus the matter in the core is very much like the ``deconfined" quark matter arrived at by  \cite{evidence}.  But there are basic differences between our system and what's described in \cite{evidence}. First of all,  in our theory, conformality is broken, though perhaps only slightly at high density, in the system.  There can also be fluctuations around $v_{pcs}^2/c^2=1/3$ coming from the effects by the anomalous dimension $\beta^\prime$. This effect can be actually seen in Fig.~\ref{fig:EoS} where  our prediction for $P/\epsilon$ is compared with the ``conformality band" obtained by the SV interpolation method~\cite{evidence}. We see that it is close to, and parallel with, the conformality band, but most significantly, it lies outside of this band.
Most important of all, the  half-skyrmion fermion is {\it not} deconfined. It is a quasiparticle of fractional baryon charge, neither purely baryonic nor purely quarkonic. In fact it can be anyonic lying on a (2+1) dimensional sheet~\cite{D}. Our conjecture is that  it represents the manifestation of an emergent scale symmetry pervading at low density -- as in $g_A^{\rm Landau}$--  and  at high density  in the vicinity of DLFP -- as in $g_A^{DL}$.
\subsubsection{Cheshire-Catism}

We will now argue that what was deemed to be the ``deconfined quarks" in \cite{evidence} could be not so different from what is found in the PC model. Modulo different approximations made in their different starting points,  top-down~\cite{evidence} and bottom-up (this paper), they could very well be coming close to the same physics. For instance, it seems plausible that the approach to the baryon-quark continuity that exploits explicit quark degrees of freedom, suitably hybridized with baryons at density $\sim 3n_0$, such as for instance the ``quarkyonic" structure on the Fermi sea~\cite{quarkyonic},  could be made to come close to the topology-change mechanism exploited in our approach by ``fine-tuning"  (arbitrary)   parameters of the baryon-quark hybrid construction.  Such ``fine-tuning"  is not in fact devoid of physics because it is tantamount to taking into account of strong correlations involved in nuclear dynamics inaccessible by controlled (e.g., perturbative) QCD tools. What may be crucially involved here are nuclear short-range correlations with repulsive interactions due to exclusion principle, i.e., ``excluded volume"~\cite{excluded-volume}, entering at density $\sim (2-3) n_0$. Such effect can be present in the constituent quark model between two nucleons~\cite{hatsuda-lee}  and in bound half-skyrmions~\cite{PKLMR}\footnote{It is known that the constituent quark model and the skymion model are equivalent at the large $N_c$ limit.}.

\begin{figure}[bhtp]
\begin{center}
\vskip 0.3cm
\includegraphics[width=0.47\textwidth]{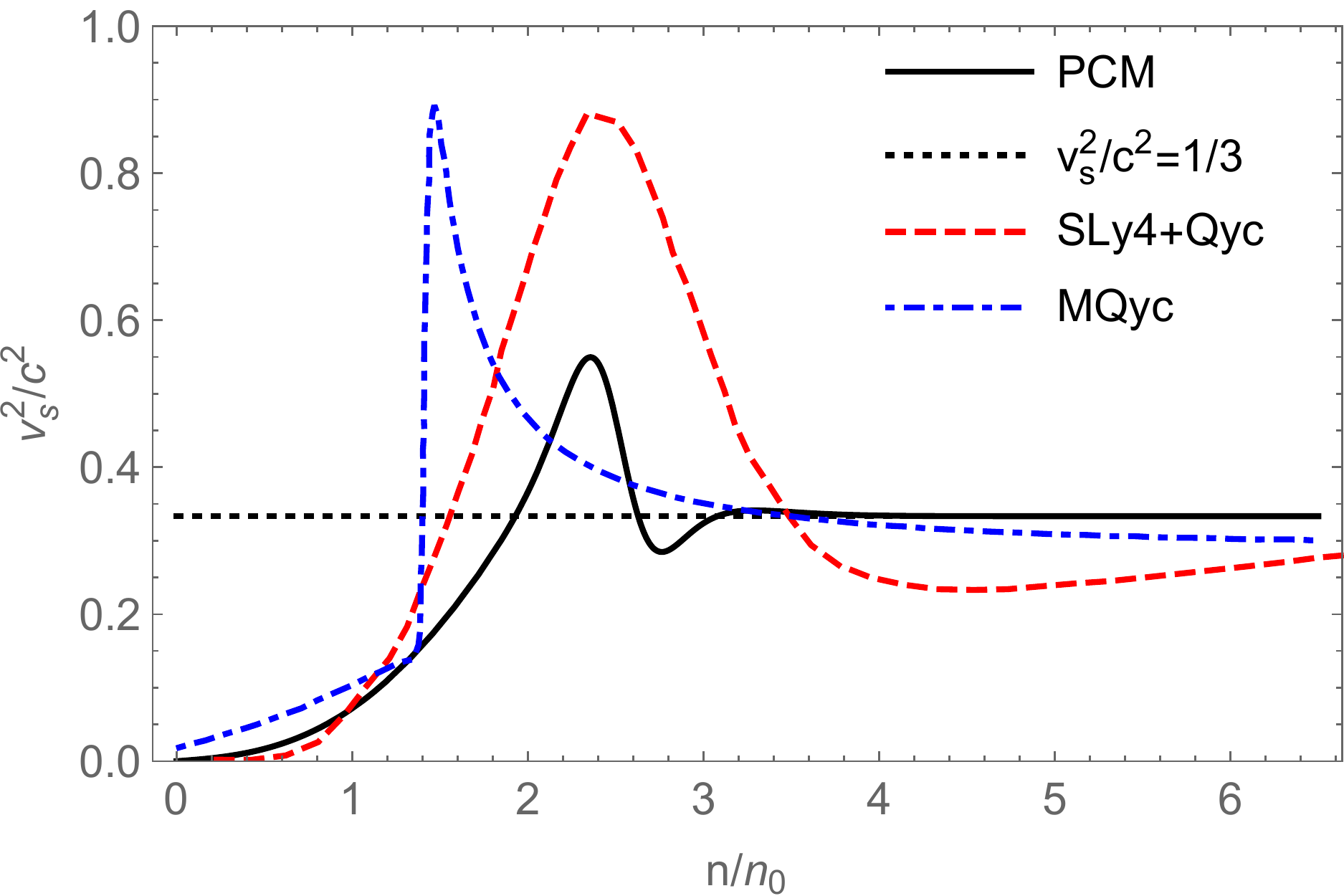}
\end{center}
\vskip -0.5cm
\caption{ Predictions of various models that could be fine-tuned to pseudo-conformality \`a la Cheshire-Catism.}
\label{othermodels}
\end{figure}

To gain a rough idea, we plot in Fig.~\ref{othermodels} a few results of $v_s$ in two different classes of hybrid baryon-quark models:  two recent results in quarkyonic model labeled ``MQyc" ~\cite{lattimer-quarkyonic} and ``SLy4+Qyc"~\cite{margueron}  and one labeled as  ``MF-Quark"~\cite{kapusta} that  suitably matches  a Walecka-type mean-field model at low density to perturbative quark model at two-loop order at high density. We interpret both -- and other similar models -- as a ``microscopic" rendition of our topology-change mechanism -- which is ``macroscopic" -- for the baryon-quark continuity.  Among those found in the literature we pick only three hybrid models to just illustrate our argument. The physics of our PCM from  $\sim n_0$ to $\sim$ the putative ``baryon-quark" changeover density in terms of the cusp structure given in the skyrmion crystal description could very well be an oversimplification as one can see in various models in terms of explicit quarks. As we will argue below, it should make little physics sense to focus on the changeover region in close detail. Nonetheless all quarkyonic-type models seem to have qualitatively similar ``bump" behavior, some broader and some sharper, as the PCM in $v_s^2/c^2$ at the changeover regime and also in their proximity  to our  pseudo-conformal velocity at $\gsim 3n_0$. As  for the hybrid MF-quark-type model,  it also exhibits similar bumps for certain parameter choices, but they seem to have wrong asymptotic behaviors. The one we picked marked ``MF-Quark" does not possess the bump but exhibits a more consistent high density behavior going toward the pseudo-conformal velocity.  In both cases,  we suggest that  given the complicated changeover process more or less totally un-controlled by trustful theoretical tools (such as lattice QCD), there could very well be a number of free parameters available in their constructions that could be adjusted to bring $v_s$ come  closer to the PCM structure at density $\gsim 3 n_0$ {\it without spoiling} other global star properties. We consider this highly plausible given the striking difference in $v_s$ observed in our PCM between $n_{\rm VM}\sim 6n_0$ and $\sim 25n_0$ without affecting appreciably  all other global star properties.~\footnote{{It would be interesting if the effects of  $n_{\rm VM}$ at low density $\sim 6n_0$ vs. at asymptotic density $\gsim 25 n_0$ could be diagnosed by gravity waves, i.e.,  the waveforms of the gravitational waves emitted from the neutron star mergers sensitive to the EOS of dense matter~\cite{Yang:2020awu}.}}    This observation leads us to suggest that {\it the underlying mechanism for such dynamical models  is governed by  approximate hidden symmetries.}   At what density within the range $\sim (2-4) n_0$ such a mechanism sets in must involve details of what's put in the changeover region. 

\section{Conclusion} 
We have suggested that quark-like degrees of freedom, if observed in the interior of massive neutron stars, can be interpreted as confined quasi-particles of fractional baryon charges in consistency with hadron-quark continuity. Such fractionally-charged objects are inevitable by topology at high densities~\cite{D}. The mechanism in action is the emergence of conformal (or scale) symmetry in interplay with the hidden local symmetry, coming not necessarily from the QCD proper, but from strongly-correlated nuclear interactions, which could permeate in baryonic matter from low density to high density. In this scheme, true deconfinement leading to genuinely deconfined quarks is to set in, as mentioned above, at much higher densities, say, $\gsim 25 n_0$, possibly with the phase transition from a Higgs mode to a topological mode conjectured \`a la Seiberg-duality between the vector mesons of HLS and the gluons of QCD~\cite{kanetal}. The bottom line is that it could be the Cheshire Catism~\cite{CC} that produces the hadron-quark continuity, a smooth crossover, driven by the hidden symmetries, coming from topology. 

The scenario that we have proposed  looks at first sight drastically different from other scenarios found in the literature although it does share  similar global star properties  with them other than the sound speed and the core structure. The crucial issue then is what can be a ``smoking-gun" signal that rules out the putative hadron-quark continuity based on topology that is our scenario as opposed to that based on quarkonic structure such as  the quarkyonic model.  As shown in this note, it is not the property of the core of stars that will be the judge since what may be identified as deconfined quarks in the hybrid hadron-quark model can be fractionalized stuffs that are not deconfined quarks.  What distinguishes the pseudo-conformal structure from others is the permeation of hidden symmetries in nuclear medium as illustrated in the precocious onset of the PC sound speed and in the quenching mechanism of $g_A$ in finite nuclei and at the dilaton limit fixed point density. It is totally unclear what signal could distinguish the pseudo-conformal sound velocity in the interior of the star as predicted by the PCM from the speed that exceeds $v_s^2/c^2\approx 1/3$ as in RMF models or possibly also in quarkyonic models. 

From the QCD point of view, the crossover region from hadrons to quarks at the relevant range of densities, $\sim (2-4)n_0$, is most likely the worst controlled region and cannot be used to gauge the validity of the models involved. In fact this was already apparent in the PCM~\cite{MR-PPNP}  when the crossover density $n_{1/2}$ is taken to be the upper bound, $n_{1/2}=4n_0$:  There was a spike in $v_s$ that exceeds the causality bound. As noted above, perhaps the waveforms of the gravity waves from merging neutron stars could carry information on this crossover region~\cite{Yang:2020awu}.

\sect{Acknowledgments}
The work of YLM was supported in part by the National Science Foundation of China (NSFC) under Grant No. 11875147 and 11475071.

\end{document}